\documentclass[pss]{wiley2sp} 
\usepackage{amsmath}
\usepackage{bm}             

\newcommand\scalemath[2]{\scalebox{#1}{\mbox{\ensuremath{\displaystyle #2}}}}

\begin{document}

\title{Zero-energy vortices in Dirac materials}

\author{%
  C.~A.~Downing\textsuperscript{\textsf{\bfseries 1}} and
  M.~E.~Portnoi\textsuperscript{\Ast,\textsf{\bfseries 2},\textsf{\bfseries 3}}}
 
\mail{e-mail
  \textsf{m.e.portnoi@exeter.ac.uk}}

\institute{%
  \textsuperscript{1}\,Departamento de F\'{i}sica de la Materia Condensada,
CSIC-Universidad de Zaragoza, Zaragoza E-50009, Spain\\
  \textsuperscript{2}\,School of Physics, University of Exeter, Stocker Road, Exeter EX4 4QL, United Kingdom\\
  \textsuperscript{3}\,ITMO University, St. Petersburg 197101, Russia}


\abstract{\bf%

In this brief review, we survey the problem of electrostatic confinement of massless Dirac particles, via a number of exactly solvable one- and two-body models. By considering bound states at zero energy, we present a route to obtain truly discrete states of massless Dirac particles in scalar potentials, circumventing the celebrated Klein tunnelling phenomenon. We also show how the coupling of two ultrarelativistic particles can arise, and discuss its implications for cutting-edge experiments with two-dimensional Dirac materials. Finally, we report an analytic solution of the two-body Dirac-Kepler problem, which may be envisaged to bring about a deeper understanding of critical charge and atomic collapse in mesoscopic Dirac systems.
 }

\maketitle   

\section{Introduction}
\label{sec:intro}

After the turn of the 21st century, the discovery of Dirac materials led to the Dirac equation becoming a cornerstone of modern mesoscopic physics \cite{Thaller1956}, \cite{Wehling2014}, \cite{Vafek2014}, \cite{Goerbig2014}. Starting with the experimental realization of graphene \cite{Katsnelson2012}, and followed swiftly after by the unearthing of topological insulators \cite{Shen2012}, the study of massless Dirac fermions became increasingly important from both fundamental and applied perspectives.

One of the most famous phenomena associated with Dirac particles is Klein tunneling \cite{Klein1929}, an effect in relativistic quantum mechanics whereby a particle impinging on a barrier tunnels into the Dirac sea of antiparticles, and in doing so heavily suppresses the chances of reflection \cite{Calogeracos1999}. In Dirac materials, there is a direct analogy when electrons inter-band tunnel into hole states \cite{Katsnelson2006}. The most famous result in this area is the perfect transmission of Dirac electron normally incident on a high potential barrier \cite{Fuchs2011} \cite{Zalipaev2012}, which has been observed experimentally in graphene \cite{Young2009}.

The appearance of Klein tunneling in Dirac materials raises the question of how to confine massless charge carriers. This problem is of crucial importance for future electronic devices which require the complete control of these elusive particles - most clearly for transistors, which require well-defined on-off digital logic. Various ideas have been proposed to induce bound states of massless Dirac fermions, including the application of magnetic fields, strain engineering, chemical doping and the opening up of a band gap, as reviewed in Ref.~\cite{Rozhkov2011}. However, these methods in some sense blunt the key attribute which made Dirac materials so attractive initially, namely their high electron mobility. Therefore, a hope remains that electrostatic confinement of ultrarelativistic particles is possible with negligible cost to their superlative transport properties.

In this brief review, we recount the efforts made to trap Dirac particles electrostatically. In particular, we show how considerations of zero-energy single particle states in scalar potentials allows one to indeed uncover bound states, albeit with some restrictions \cite{Hartmann2010}, \cite{Calvo2011}, \cite{Stone2012}, \cite{Ioffe2018}. We go on to describe the ultra-relativistic two body problem \cite{Sabio2010}, detail how the coupling of two massless Dirac fermions is indeed possible, and outline its potential consequences \cite{Portnoi2017}. We also comment on atomic collapse in Dirac materials \cite{Shytov2007b}, and present an analytical solution for the two-body Dirac-Kepler problem, which has implications for ongoing research into the full many-body problem \cite{Kotov2012}. 

The rest of this work is organized as follows. In Sec.~\ref{sec:bound}, we provide the continuum theory behind the absence of bound states in Dirac materials. We highlight how electrostatic confinement is indeed possible at zero-energy in Sec.~\ref{sec:zeroes}, and discuss the latest experiential work in the area in Sec.~\ref{sec:expt}. We introduce the formalism of the two-body problem in Sec.~\ref{sec:twobody}, where we provide the exact solution of the Dirac-Kepler problem and review the possible nontrivial pairing at zero-energy. In Sec.~\ref{sec:other} we outline competing proposals to create bound states in Dirac materials, and Sec.~\ref{sec:zeromodes} is devoted to a brief survey of zero-energy states in wider physics research. Finally, we draw some conclusions in Sec.~\ref{sec:conc}, and sketch out some perspectives for this direction of research topic.


\section{On bound states with the massless Dirac equation}
\label{sec:bound}

Here we examine the possibility to obtain bound states from the addition of scalar potentials to the massless Dirac equation. Our modus operandi is asymptotic analysis of the Dirac equation for arbitrary potential wells, where we search for square-integrable solutions associated with bound states [Sec.~\ref{sec:long}]. We treat the Coulomb problem separately due to its long-range nature [Sec.~\ref{sec:onecou}].

The single particle 2D Dirac-Weyl Hamiltonian reads \cite{Katsnelson2012}, \cite{Shen2012}
\begin{equation}
\label{eq1}
H = v_{\mathrm F} \boldsymbol \sigma \cdot \boldsymbol{\hat p} + V(r),
\end{equation}
where $v_{\mathrm F}$ is the Fermi velocity of the material, the momenta $\boldsymbol{\hat p} = (\hat p_x, \hat p_y)$, $\boldsymbol \sigma = (\sigma_x, \sigma_y, \sigma_z)$ are Pauli's spin matrices and $V(r)$ is a scalar potential. We shall act on the Hamiltonian~\eqref{eq1} with a two-component spinor wavefunction in the form 
\begin{equation}
\label{eq2}
\Psi(r,\theta) = \frac{e^{ \mathrm{i} m\theta} }{\sqrt{2\pi}} \left(
 \begin{array}{c}
\chi_A(r) \\ \mathrm{i} e^{ \mathrm{i} \theta}\chi_B(r)
 \end{array}
\right),
\end{equation}
where the quantum number $m=0,\pm1, \pm2, ...$ is directly related to the total angular momentum quantum number $j_z = m +1/2$. Explicitly, the spinor~\eqref{eq2} satisfies $J_z \Psi = j_z \Psi$, with $J_z = -\mathrm{i} \partial_{\theta} + \sigma_z/2$. The wavefunction~\eqref{eq2} separates the variables in the Schr\"{o}dinger equation $H \Psi = E \Psi$, leading to a pair of coupled equations for the radial components
\begin{subequations}
\label{eq3}
 \begin{align}
  \left(\partial_r + \tfrac{m+1}{r} \right) \chi_B &= [\varepsilon-U(r)] \chi_A, \label{eq4} \\
  \left(-\partial_r + \tfrac{m}{r}   \right) \chi_A &= [\varepsilon-U(r)] \chi_B, \label{eq5}
 \end{align}
\end{subequations}
where the scaled eigenvalue $\varepsilon = E/\hbar v_{\mathrm F}$ and scaled potential function $U(r) = V(r)/\hbar v_{\mathrm F}$. Decoupling the two first-order equations~\eqref{eq3} leads to the following second-order differential equation for the upper component of the radial wavefunction
\begin{equation}
\label{eq6}
\chi_A'' + \left( \tfrac{1}{r} + \tfrac{U'}{\varepsilon-U} \right) \chi_A' + \left( \left[ \varepsilon-U \right]^2 - \tfrac{m^2}{r^2} - \tfrac{m}{r} \tfrac{U'}{\varepsilon-U} \right) \chi_A = 0,
\end{equation}
here $'$ denotes taking a derivative with respect to $r$. The corresponding equation for $\chi_B$ is obtained with the replacements $A \to B$ and $m \to -(m+1)$ in Eq.~\eqref{eq6}. In the following two subsections, we consider the asymptotics of Eq.~\eqref{eq6}, as well as the special case of the Coulomb problem.


\subsection{Long-range asymptotics}
\label{sec:long}

As pointed out by Tudorovskiy and Chaplik \cite{Tudorovskiy2007}, the large-$r$ asymptotics of Eq.~\eqref{eq6} for some (faster than Coulomb) decaying potential well is exactly Bessel's differential equation 
\begin{equation}
\label{eq7}
\chi_A'' + \left( \tfrac{1}{r} \right) \chi_A' + \left( \varepsilon^2 - \tfrac{m^2}{r^2} \right) \chi_A = 0,
\end{equation}
since the terms with $U(r \to \infty)$ and $U(r\to \infty)'$ can be safely neglected. However, unlike massive particles as governed by the 2D Schr\"{o}dinger equation, the scaled energy appears here as squared ($\varepsilon^2$). Therefore the asymptotic solutions of Eq.~\eqref{eq7} are independent of the sign of the energy, and thus behave like the scattering (and not the bound state) solutions of the equivalent problem for non-relativistic particles. Explicitly, the solutions of Eq.~\eqref{eq7} are the superposition
\begin{equation}
\label{eq7b}
\chi_A = c_1 J_m(\varepsilon r) + c_2 Y_m(\varepsilon r), 
\end{equation}
in terms of the Bessel functions of the first [$J_m(z)$] and second [$Y_m(z)$] kinds of order $m$, and where $c_{1,2}$ are some constants. The solution~\eqref{eq7b} suggests a slow asymptotic decay like $\chi_{A, B} \propto 1/\sqrt{r}$, which is characteristic of scattering states. Thus the question of how to create bound states, which are defined by square-integrable wavefunctions with the typical exponential decay $\chi_A \propto e^{- r / r_0}$, is raised.


\subsection{One-particle Coulomb problem}
\label{sec:onecou}

The above asymptotic analysis does not hold for long-range potentials, although it can be shown it does not change the fact that discrete states are seemingly impossible to create.

We consider the 2D Dirac-Kepler problem \cite{Guo1991} \cite{Shytov2007b}, \cite{Shytov2007a}, \cite{Pereira2007}, \cite{Novikov2007}, \cite{Shytov2009}, \cite{Gupta2009}, \cite{Chakraborty2011} with the Coulomb potential
\begin{equation}
\label{eqCou0}
U(r) = -\frac{\alpha}{r}, 
\end{equation}
where $\alpha$ is the dimensionless strength parameter (and indeed the effective fine structure constant for 2D Dirac materials). One may find from Eqs.~\eqref{eq3} the solutions
\begin{equation}
\label{eqCou1}
\chi_{A, B} = r^{\gamma_m-1/2} e^{-\mathrm{i} \varepsilon r} f_{A, B}( \varepsilon r), 
\end{equation}
where the somewhat complicated functions $f_{A, B}( \varepsilon r)$ are derived explicitly in Ref~\cite{Shytov2007a}. In Eq.~\eqref{eqCou1}, the `atomic collapse' parameter
\begin{equation}
\label{eqCou2}
\gamma_m = \sqrt{ {\alpha^{\mathrm{c}}_m}^2 -\alpha^2}, 
\end{equation}
defines two regimes of the problem: sub-critical ($\alpha<\alpha^{\mathrm{c}}_m$) with real $\gamma_m$; and super-critical ($\alpha>\alpha^{\mathrm{c}}_m$) with imaginary $\gamma_m$. The critical strength of the Coulomb potential is given by the $m$-dependent quantity
\begin{equation}
\label{eqCouweww2}
\alpha^{\mathrm{c}}_m = |m+1/2|,
\end{equation}
with important threshold value $\alpha^{\mathrm{c}}_{0} = 1/2$. In the super-critical regime, there are pathological oscillations in the wavefunction at short range ($r \to 0$), mimicking the celebrated atomic collapse phenomena of an electron falling into the nucleus \cite{Pomeranchuk1945}, \cite{Zeldovich1972}. Furthermore, so-called quasi-bound states may also appear \cite{Shytov2007b}, however there are no truly bound states with square-integrable wavefunctions, as can be seen from the $\chi_{A, B} \propto 1/\sqrt{r}$ asymptotic decay in Eq.~\eqref{eqCou1} with $r \to \infty$.

The hallmarks of atomic collapse are believed to have been seen in two landmark experiments. In Ref.~\cite{Wang2013}, the supercritical Coulomb regime was reached by pushing clusters of calcium dimers together on the surface of gated graphene devices, with the aid of a scanning tunneling microscope (STM). STM spectroscopy was then used to probe the atomic collapse-like states as a function of the Fermi level, which is tunable via electrostatic gating. In Ref.~\cite{Mao2016}, the authors exploit a single-atom vacancy to create a stable supercritical charge. Using an STM tip to apply voltage pulses to the vacancy, the positive charge can be increased so that nearby conduction electrons experience the supercritical phase. Resultant measurements through STM microscopy and Landau level spectroscopy suggest the signatures of atomic-collapse states in the density of states.

It should also be mentioned that results of the atomic cluster experiment of Ref.~\cite{Wang2013} leave some option questions. For example, the peak in the density of states for critically charged clusters appears precisely at the Dirac point, rather than a few electron volts below it, and with the further increase of charge it even moves in the `wrong' direction above the Dirac point. In addition, the density distribution in the `collapsed' states is crater-like and covers tens of lattice constants, which are all signatures of a fully-confined zero-energy state as discussed in the next section.

The atomic collapse observed in graphene is not directly analogous to the standard textbook case of atomic collapse \cite{Berestetsky1982}, since the particles are massless and hence there is no bandgap. However, the characteristic feature of bound states diving into the continuum can be seen with the massive Dirac particles, as found in certain carbon nanotubes for example \cite{Downing2014}.


\section{Zero energy states of the massless Dirac equation}
\label{sec:zeroes}

The no-go theorem of the Sec.~\ref{sec:long} in fact requires an important caveat, which we detail here. This stipulation opens up a route to create bound states electrostatically, a feat against popular and widely held beliefs.

The impossibility of supporting bound states of massless Dirac fermions was based on an asymptotic analysis of Eq.~\eqref{eq7}. However, so as to avoid a reduction to the scattering states of Eq.~\eqref{eq7b}, one may consider the special case of zero-energy states ($\varepsilon=0$), which were previously implicitly ignored. Then the asymptotic solution of Eq.~\eqref{eq7} is instead $\chi_A \propto 1/r^{|m|}$, such that bound states may indeed form for certain values of angular momentum as indexed by $m$ \cite{Downing2011}, \cite{Downing2015}. The resultant algebraically decaying wavefunctions are a characteristic of this problem, in stark contrast to the usual exponential decay of non-relativistic problems. Furthermore, we note that the sign of the potential is not important for the quantization condition, since the potential enters Eq.~\eqref{eq7} as either squared or as a logarithmic derivative for $\varepsilon=0$.

In what follows, we investigate two simple models which elucidate the features of such zero energy bound states [Secs.~\ref{sec:squarecircle}~and~\ref{sec:Lorentzian}], before analyzing the zero-energy Coulomb problem [Sec.~\ref{sec:shytov}].


\subsection{Finite circular well}
\label{sec:squarecircle}

Let us start by considering the simplest possible model, of bound states in the finite circular well \cite{Hewageegana2008}, \cite{Bardarson2009}, \cite{Ahmadi2015}
\begin{equation}
\label{eq8}
U(r) = -U_0 \Theta(a-r),
\end{equation}
where $U_0$ and $a$ are the strength (in units of inverse length) and spatial range of the well respectively, and $\Theta(z)$ is the Heaviside step function. 

Solving the coupled equations~\eqref{eq3} inside the potential well ($r<a$) yields
\begin{equation}
\label{eq9}
\left(
 \begin{array}{c}
\chi_A^{\mathrm{in}} \\ \chi_B^{\mathrm{in}}
 \end{array}
\right) = \frac{C_{nm}}{a} \left(
 \begin{array}{c}
J_m(U_0 r) \\ J_{m+1} (U_0 r)
 \end{array}
\right),
\end{equation}
where $C_{nm}$ is a normalization constant to be determined. Outside the potential well ($r>a$), one finds
\begin{equation}
\label{eq10}
\left(
 \begin{array}{c}
\chi_A^{\mathrm{out}} \\ \chi_B^{\mathrm{out}}
 \end{array}
\right) = \frac{C_{nm}}{a} \left(
 \begin{array}{c}
0 \\ J_{m+1} (U_0 a) \left( \frac{a}{r} \right)^{m+1}
 \end{array}
\right), m = 0, 1,...
\end{equation}
The requirement of continuity of both of the radial wavefunction components leads to the condition $J_{m}(U_0a) = 0$, such that bound states appear at
\begin{equation}
\label{eq11}
U_0 a = \alpha_{m, n},
\end{equation}
where $\alpha_{m, n}$ is the $n$-th zero of the Bessel function of the first kind of order $m$. Therefore, there are a countably infinite number of zero-energy bound states, which occur at certain critical values of the potential strength. Crucially, the states with $m = 0$ are extended states (appearing at $U_0 a \simeq 2.41, 5.52, 8.65...$), since they are marginally non-integrable due to their slow algebraic decay [from Eq.~\eqref{eq10}, $\chi_B \propto 1/r$]. The first square-integrable bound state is for $m = 1$ (where $\chi_B \propto 1/r^2$), at the threshold potential strength $U_0 a \simeq 3.83$. Further bound states successively appear with increasing the potential strength $U_0 a$. All of the bound-state supporting values of Eq.~\eqref{eq11} are distinct, since the Bessel function of the first kind of integer order $J_{m+n}(z)$ does not have common zeros with $J_{m}(z)$ as follows from Bourget's hypothesis \cite{Watson1966}.

Most notably, Eq.~\eqref{eq10} displays a complete suppression of probability density on the upper wavefunction component ($\chi_A$), in a direct manifestation of the chirality of the system. When the model describes electrons in a honeycomb lattice, like in graphene, this means the electron density is only on the B sublattice of the honeycomb structure.

The analysis leading to Eq.~\eqref{eq10} may be extended for negative values of $m$, yielding the solution
\begin{equation}
\label{eq12}
\left(
 \begin{array}{c}
\chi_A^{\mathrm{out}} \\ \chi_B^{\mathrm{out}}
 \end{array}
\right) = \frac{C_{nm}}{a} \left(
 \begin{array}{c}
J_{m} (U_0 a) \left( \frac{r}{a} \right)^{m} \\ 0
 \end{array}
\right), m = -1, -2,...
\end{equation}
Now the outer radial wavefunction demonstrates the absence of probability density on the lower wavefunction component ($\chi_B$), due to its zero-eigenvalue condition $J_{m+1}(U_0a) = 0$, or $U_0 a = \alpha_{m+1, n}$. The solution~\eqref{eq12} shows that states with $m=-1$ are extended and those with $m\le-2$ are bound states. All of these features highlight the intrinsic $A \to B$ and $m \to -(m+1)$ symmetry of the system, as follows from the governing Eq.~\eqref{eq3}.

It is instructive to now consider a more realistic, smooth potential well, which has the added benefit of an analytic expression for the supported zero-energy states.


\subsection{Lorentzian well}
\label{sec:Lorentzian}

We shall now consider fully-confined zero-modes trapped in the smooth confining potential \cite{Downing2011} \cite{Downing2017} 
\begin{equation}
\label{toy1}
U(r) = \frac{-U_0}{1+r^2/d^2},
\end{equation}
where $U_0$ and $d$ parametrize the strength (in units of inverse length) and spatial extent of the potential respectively.

We wish to solve the coupled Eq.~\eqref{eq3} by initially finding just the upper radial wavefunction component $\chi_A$. A short-range analysis of Eq.~\eqref{eq6} suggests that at $r \to 0$ our solution should be of the form $\chi_A \sim r^{|m|}$; whilst a similar long-range analysis suggests a decay like $\chi_A \sim r^{|m|  - 2 p_m}$ as $r \to \infty$. Here we have introduced the parameter
\begin{equation}
\label{toy2}
p_m = \tfrac{1}{2} \left( 1 + |m| + |1+m| \right).
\end{equation}
We switch to a new variable $\xi = (r/d)^2$, and try an appropriate ansatz solution of Eq.~\eqref{eq6} in the form  
\begin{equation}
\label{toy3}
\chi_A = \frac{\xi^{\tfrac{|m|}{2}}}{(1+\xi)^{p_m}} w(\xi),
\end{equation}
where $w(\xi)$ is an unknown function, having no bearing on the small or large asymptotics of the solution. We find from Eq.~\eqref{eq6} the following equation for $w(\xi)$ 
\begin{align}
\label{toy4}
 &\xi(1+\xi)^2 w''(\xi) \nonumber \\
 &+ (1+\xi) \left[ 1 + |m| + (2+|m|-2 p_m) \xi \right] w'(\xi) \nonumber \\
 &+ \left[ (\tfrac{U_0 d}{2})^2 - p_m^2 \right] w(\xi) = 0,
\end{align}
The solution of Eq.~\eqref{toy4} can be given in terms of the Gauss hypergeometric function $\,_2F_1\left( a, b; c; z\right)$ as
\begin{equation}
\label{toy5}
 w(\xi) =  \,_2F_1\left(  p_m + \tfrac{1}{2} U_0 d  , p_m - \tfrac{1}{2} U_0 d ;  1 + |m| ; \tfrac{\xi}{\xi+1} \right).
\end{equation}
Thus we have found, from Eq.~\eqref{toy3} with Eq.~\eqref{toy5}, the explicit form of $\chi_A$, while $\chi_B$ is readily obtained from Eq.~\eqref{eq5}. It is implicit that the power series in Eq.~\eqref{toy5} must be terminated, so as to satisfy the required conditions on its limiting behavior determined at the outset [given above Eq.~\eqref{toy2}]. Given this termination, the radial asymptotics are
\begin{equation}
\label{toy121}
\lim_{r \to \infty}
\left(
 \begin{array}{c}
\chi_A\\ \chi_B
 \end{array}
\right) \propto \frac{1}{r} \left(
 \begin{array}{c}
\frac{1}{r^|1+m|} \\ \frac{1}{r^|m|}
 \end{array}
\right), 
\end{equation}
which highlights the algebraically decaying, but marginally non-square-integrable, extended states with $m=\{0, -1\}$ and the otherwise bound states with $m\ne\{0, -1\}$, as was the case of the finite circular well of the previous Subsection~\ref{sec:squarecircle}.

We obtain the conditions for bound states by assigning the second argument in Eq.~\eqref{toy5} to be a non-positive integer, which results in
\begin{equation}
\label{toy6}
 U_0 d = 2 \left( n + p_m \right), \quad n = 0, 1, 2 ...
\end{equation}
where $p_m$ is defined in Eq.~\eqref{toy2}. Therefore a staircase of bound states is formed with increasing potential strength $U_0 d$, starting from the threshold of $U_0 d = 4$, since $U_0 d = 2$ only supports extended states with $m=0$ or $m=-1$. Clearly, all of the modes of Eq.~\eqref{toy6} appear at even integer values of $U_0 d = 2 N$, where $N=1, 2, 3, ...$ and possess an accidental degeneracy of $2N$. For example, when $U_0 d = 2$ there is a two-fold degeneracy with the states $(n, m) = (0, 0)$ and $(0, -1)$, while for $U_0 d = 4$ there is a four-fold degeneracy, encompassing $(n, m) = (0, 1)$, $(1, 0)$, $(0, -2)$ and $(1, -2)$.

Our analyses of the toy models in this section allows us to gather together some general conclusions. In order to have bound states, we require: zero-energy states ($\varepsilon = 0$); a scalar potential $U(r)$ decaying faster than $1/r$, a critical strength $U_0$ and spatial extent $d$ of the well, and a large enough angular momentum, or vorticity, $m \ne \{0, -1\}$. Therefore, we term these exotic bound states `zero-energy vortices'.
 

\subsection{Coulomb problem}
\label{sec:shytov}

It is also worth considering zero-energy states with the Coulomb potential. At finite energy ($\varepsilon \ne 0$) the exact solution is in terms of  non-square-integrable eigenfunctions, leading to the absence of bound states as was discussed in Sec~\ref{sec:onecou}. However, the special case of zero-energy states ($\varepsilon = 0$) which interests us were not considered before. Indeed, zero-energy states were implicitly excluded by the variable choice $\varepsilon r$ in Eq.~\eqref{eqCou1}.

We study the regularized Coulomb potential
\begin{equation}
\label{cou1}
U(r) = 
  \begin{cases} 
   -\frac{\alpha}{R}, &  r \le R, \\
   -\frac{\alpha}{r}, & r > R,
  \end{cases}
\end{equation}
with the (dimensionless) strength $\alpha$ and length scale $R$. Inside the cutoff region ($r < R$) we find
\begin{equation}
\label{cou2}
\left(
 \begin{array}{c}
\chi_A^{\mathrm{in}} \\ \chi_B^{\mathrm{in}}
 \end{array}
\right) = \frac{C}{R} \left(
 \begin{array}{c}
J_m(\alpha r/R) \\ J_{m+1} (\alpha r/R)
 \end{array}
\right),
\end{equation}
where $C$ is the normalization constant. Outside the cutoff region ($r < R$), we find
\begin{equation}
\label{cou3}
\left(
 \begin{array}{c}
\chi_A^{\mathrm{out}} \\ \chi_B^{\mathrm{out}}
 \end{array}
\right) = \frac{C }{R} J_m(\alpha) \left(
 \begin{array}{c}
1 \\ \frac{ m + \gamma_m + 1/2 }{\alpha}
 \end{array}
\right) \left( \frac{R}{r} \right)^{\gamma_m+1/2},
\end{equation}
where we have used the atomic collapse parameter defined in Eq.~\eqref{eqCou2}. In order to describe a sufficiently fast decaying wavefunction, one requires $m(m+1)>\alpha^2$. The wavefunction continuity condition on $\chi_B$ suggests that bound states should appear at certain values of $\alpha$, which satisfy
\begin{equation}
\label{cou4}
 \alpha J_{m+1}(\alpha) = \left( m + \gamma_m + 1/2 \right) J_m(\alpha),
\end{equation}
which is notably $R$ independent. However, this transcendental equation~\eqref{cou4} has no solutions, as one can readily see graphically. Therefore we conclude that even at zero-energy, long-range potentials are not able to support bound states.

We should mention that the pure Coulomb potential is of questionable relevance for realistic low-dimensional materials since it does not account for dielectric screening, which leads to its modification as discussed by Rytova and Keldysh for thin films \cite{Rytova1967} \cite{Keldysh1979} and by Cudazzo and co-workers for 2D materials \cite{Cudazzo2011}. Indeed, the problem of zero energy states in screened potentials is a particularly fruitful avenue for further research \cite{Wang2018}.


\subsection{Adding a magnetic flux tube}
\label{sec:flux}

The addition of a magnetic flux tube to the Hamiltonian~\eqref{eq1} via the vector potential $\boldsymbol{A} = (0, \hbar f / e r, 0)$, where $f=\Phi/\Phi_0$ is the number of flux quanta $\Phi_0 = h/e$, provides a useful extra degree of freedom \cite{Heinl2013}. After introducing a generalized momentum $\boldsymbol{\hat p} \to \boldsymbol{\hat p} + e \boldsymbol{A}$ to incorporate such a vector potential, one obtains same coupled equations as without a flux tube [c.f.~Eqs.~\ref{eq3}] after the transformation $m \to \tilde{m} = m + f$. Therefore, bound states are no longer determined by just the angular momentum quantum number, with the number of flux quanta inside the flux tube giving rise to a degree of tunability which may be exploited in future experiments.


\subsection{Supersymmetric solutions}
\label{sec:super}

The experimental breakthroughs in realizing Dirac materials have led to enormous theoretical activity regarding exact solutions of Dirac-like equations. One neat method to obtain bound states for a variety of scalar potentials is through the techniques of supersymmetric quantum mechanics \cite{Cooper2001}, \cite{Gangopadhyaya2010}, \cite{Hirshfeld2012} as applied to the Dirac equation \cite{Ho2014}, \cite{Ghosh2016}. Further analytic solutions have been generated by exploiting the relation between solitonic solutions of the Kortweg-de Vries equation and the Dirac equation \cite{Ho2015}.


\section{Experimental signatures of confinement}
\label{sec:expt}

The principle signature of bound or quasi-bound states in Dirac materials is through the local density of states data obtained by STM measurements \cite{Schneider2014}, \cite{Nguyen2017}. As mentioned previously, the results of the atomic cluster experiments of Ref.~\cite{Wang2013} can also be explained by full-confined zero-energy states. In particular, the movement of the apparent maximum in the density of states above the Dirac point for supercritical clusters can be explained by screening of the clusters by free electrons to precisely the critical potential strength value, producing the combined picture of a zero-energy states peak plus an additional screening electron density.

A landmark experiment on confinement in graphene focused on whispering-gallery modes \cite{Zhao2015}, so-called because of how sound waves in the circular chamber walls of St. Paul's Cathedral in London are confined by the principle of reflection. Resonators were made by circular pn junctions created by a scanning tunneling probe and Klein scattering of the electron waves at the edges of the formed cavity led to resonances in the local density of states.

In Ref.~\cite{Lee2016}, the authors used STM to map the electronic structure of massless Dirac fermions confined by circular graphene pn junctions,  allowing for the observation of the energy levels and nodal patterns inside the quantum dot. STM was also used in Ref.~\cite{Gutierrez2016} to probe the transmission of Dirac electrons through a sharp circular potential well defined by substrate engineering. Later experimental work on confinement in circular pn junctions created by the tip-induced charge, has seen signatures of both atomic collapse states and whispering-gallery modes \cite{Jiang2017}. Notably, the theoretical model for the tip potential was based on an infinite harmonic well which is known to not have bound states, and the truly bound states at zero-energy were overlooked. Therefore a consistent treatment of whispering gallery modes with a potential reaching a plateau at long range, rather than an infinity, is ripe for investigation.

Monolayer graphene was also studied in Ref.~\cite{Freitag2016}, where the authors studied the interplay of a homogeneous magnetic field and electrostatic confinement. An STM tip was used to induce a confining potential in the Landau gaps of bulk graphene, thus circumventing Klein tunneling. A magnetic field was also exploited in Ref.~\cite{Ghahari2017}, where a dramatic increase in the energy of states in circular graphene p-n junction resonators were observed when a small critical magnetic field was reached, in a manifestation of the topological behavior of Dirac particles.

In Ref.~\cite{Bai2018}, the authors realized nanoscale p-n junctions with atomically sharp boundaries in graphene monolayers by creating a monolayer vacancy island of copper surface, and they saw quasibound states with a finite lifetime in the formed graphene quantum dots. The realization of truly bound states in nanoscale graphene quantum dots was recently reported in \cite{Qiao2017}, where the quantum dots were electronically isolated by boundaries generated by strong coupling between the graphene and the substrate. The experimental quest for the trapping and controllable manipulation of electronic modes in massless Dirac materials remains an active and ongoing pursuit.


\section{Two-body problem}
\label{sec:twobody}

We now consider the simplest few-body problem, that of two interacting Dirac particles, in order to gain some insight into the many-body problem. Our aim is to understand the nature of the coupling between massless particles, considering the difficulties for confinement alluded to in the previous sections.

The ultra-relativistic two-body problem in 2D has been studied thoroughly in recent years, mainly from matrix Hamiltonian approaches \cite{Sabio2010}, \cite{Lee2012}, \cite{Sablikov2017} but also from the Bethe-Salpeter equation \cite{Gamayun2009}, \cite{Wang2011}. The analogous problem in one spatial dimension has also been extensively studied, for example in Refs.~\cite{Hartmann2011},~\cite{Ratnikov2012},~and~\cite{Hartmann2017}. 

A variety of exotic phenomena have been predicted at the few-body level, including: the superconducting coupling of electrons in graphene due to phonons \cite{Lozovik2010}; metastable electron-electron states \cite{Marnham2015}, \cite{Marnham2015b} and exciton-hole states \cite{Mahmoodian2012}, \cite{Mahmoodian2013} due to trigonal warping; the formation of excitons in double layer graphene \cite{Berman2013}; two-particle scattering \cite{Gaul2014}; and two-body bound states in gapped graphene \cite{Martino2017}. Furthermore, it is known that the self-trapping of Wannier-Mott excitons in polar crystals occurs at zero-energy, and the excitons experience a short-range self-trapping potential (decaying like $1/r^4$) due to electron-hole charge compensation \cite{Kusmartsev1983}.


\subsection{Formalism}
\label{sec:formalism}
The two-particle 2D Dirac-Weyl Hamiltonian reads \cite{Sabio2010}
\begin{equation}
\label{eqtwo1}
H = v_{\mathrm F} \boldsymbol \sigma \cdot \boldsymbol{\hat p}_1 \oplus v_{\mathrm F} \boldsymbol \sigma \cdot \boldsymbol{\hat p}_2 + V,
\end{equation}
where the momenta of particle $\{1, 2\}$ is $\boldsymbol{\hat p}_{1, 2} = (\hat p_{x_{1, 2}}, \hat p_{y_{1, 2}})$, the interaction is $V = V(\sqrt{(x_1-x_2)^2+(y_1-y_2)^2})$ and $\oplus$ denotes the direct sum. Neglecting for the moment the interaction, there are four eigenvalues from the Schr\"{o}dinger equation $H \Psi = E \Psi$, which are
\begin{equation}
\label{eqtwo11212121}
E = \pm  v_{\mathrm{F}} \sqrt{p_{x_{1}}^2+p_{y_{1}}^2} \pm v_{\mathrm{F}} \sqrt{p_{x_{2}}^2+p_{y_{2}}^2}.
\end{equation}
Now with the interaction $V$ taken into account, let us move into center-of-mass [$X=(x_{1}+x_{2})/2$, $Y=(y_{1}+y_{2})/2$] and relative motion [$x=x_{1}-x_{2}$, $y=y_{1}-y_{2}$] coordinates, before transforming into polar coordinates [$(X, Y) \to (R, \Theta)$, $(x, y) \to (r, \theta)$]. We try the solution $\Psi(\mathbf{R}, \mathbf{r}) = e^{\mathrm{i} \mathbf{K} \cdot \mathbf{R}}\psi(r,\theta)$, where $\hbar \mathbf{K}$ is the center-of-mass momentum. This leads to the equation $H_K \psi = \mathrm{i} [\varepsilon-U(r)] \psi$, with
\begin{equation}
\label{eqtwo2}
H_K = 
\scalemath{0.7}{
\begin{bmatrix}
0 & - \partial_z + \frac{K}{2} e^{-\mathrm{i} \Theta_K} & \partial_z + \frac{K}{2} e^{-\mathrm{i} \Theta_K} & 0\\
- \partial_{\bar{z}} + \frac{K}{2} e^{\mathrm{i} \Theta_K} & 0 & 0 & \partial_z + \frac{K}{2} e^{-\mathrm{i} \Theta_K} \\
\partial_{\bar{z}} + \frac{K}{2} e^{\mathrm{i} \Theta_K} & 0 & 0 & -\partial_{\bar{z}} + \frac{K}{2} e^{-\mathrm{i} \Theta_K} \\
0 & \partial_{\bar{z}} + \frac{K}{2} e^{\mathrm{i} \Theta_K} & -\partial_{\bar{z}} + \frac{K}{2} e^{\mathrm{i} \Theta_K} & 0
\end{bmatrix}},
\end{equation}
where $K = \sqrt{K_x^2+K_y^2}$, $\tan (\Theta_K) = K_y/K_x$, $\partial_z = e^{-\mathrm{i}\theta}\left(\partial_r-\tfrac{\mathrm{i}}{r}\partial_{\theta}\right)$ and $\partial_{\bar{z}}$ is its conjugate. Upon separating the variables with the following radial spinor wavefunction
\begin{equation}
\label{eq2qwqwqwqw}
\psi(r,\theta) = \frac{e^{ \mathrm{i} m \theta}}{\sqrt{2\pi}} \left(
 \begin{array}{c}
e^{ -\mathrm{i} \theta} R_1(r)\\
 R_2(r)\\
 R_3(r)\\
e^{ \mathrm{i} \theta} R_4(r)
 \end{array}
\right),
\end{equation}
and changing the basis functions to $\chi_{1, 2} = R_1 \pm R_4$ and $\mathrm{i} \chi_{3, 4} =  R_2 \pm R_3$ for simplicity, we obtain a system of four equations to be solved
\begin{subequations}
\label{eqasasasas3}
 \begin{align}
  - \frac{m}{r} \chi_4 &= \frac{1}{2} [\varepsilon-U(r)] \chi_1, \label{eqa} \\
  - \partial_r \chi_4 &= \frac{1}{2} [\varepsilon-U(r)] \chi_2, \label{eqaa} \\
  0 &= \frac{1}{2} [\varepsilon-U(r)] \chi_3, \label{eqaaa} \\
  \partial_r \chi_2 - \frac{m}{r} \chi_1 + \frac{1}{r} \chi_2 &= \frac{1}{2} [\varepsilon-U(r)] \chi_4. \label{eqaaaa} 
 \end{align}
\end{subequations}
Most notably Eq.~\eqref{eqaaa} has the trivial solution $\chi_3=0$, while the remaining trio of equations can be manipuated to provide a single equation for $\chi_4$ only:
\begin{equation}
\label{eq634343}
\chi_4'' + \left( \tfrac{1}{r} + \tfrac{U'}{\varepsilon-U} \right) \chi_4' + \left( \left[ \tfrac{\varepsilon-U}{2} \right]^2 - \tfrac{m^2}{r^2} \right) \chi_4 = 0,
\end{equation}
while is analogous to the one-particle equation~\eqref{eq6}. In what follows, we seek bound state solutions of Eq.~\eqref{eq634343}, but first we investigate the Coulomb interaction in the next subsection.


\subsection{Coulomb interaction}
\label{sec:kotov}

Most famously for Cooper pairing \cite{Cooper1956}, two-body problems in condensed matter physics are known to give unique insights into the full many-body problem. The ultra-relativistic two particle problem is highly important for a range of phenomena, from critical coupling to pair condensation to spontaneous gap generation \cite{Kotov2012}. Thus far, only the exact solution for the $m=0$ state has been reported \cite{Sabio2010}, here we seek to derive the general solution for all values of $m$.

We consider the bare Coulomb potential~\eqref{eqCou0} and attempt to first find the fourth wavefunction component $\chi_4$ from Eq.~\eqref{eq634343}. An asymptotic analysis of Eq.~\eqref{eq634343} suggests the behavior $\chi_4 \propto e^{-\mathrm{i} \varepsilon r/2}$ at large-range, and $\chi_4 \propto r^{\gamma_m - 1/2}$ at short-range. Here the two-body atomic collapse parameter [c.f. Eq.~\eqref{eqCou2}] reads
\begin{equation}
\label{eqCou22323}
\gamma_m = \tfrac{1}{2} \sqrt{{\alpha^{\mathrm{c}}_m}^2-\alpha^2}.
\end{equation}
As in the single particle case of Sec.~\ref{sec:onecou}, there are two regimes of the interest: sub-critical ($\alpha<\alpha^{\mathrm{c}}_m$) and super-critical ($\alpha>\alpha^{\mathrm{c}}_m$). The critical strength of the Coulomb potential [c.f. Eq.~\eqref{eqCouweww2}] is
\begin{equation}
\label{eqCouwe23223ww2}
\alpha^{\mathrm{c}}_m = \sqrt{1 + 4 m^2},
\end{equation}
with the first few values $\alpha^{\mathrm{c}}_{0, 1, 2} \simeq \{ 1, 1.41, 2.24\}$. These supercritical values are comparable to those reported in the literature by other means. For example, by simulating the phase diagram of graphene as a function of the Coulomb coupling between quasiparticles via Monte Carlo calculations, the critical couplings $\alpha^{\mathrm{c}}_{0} = 1.11 \pm 0.06$ were found \cite{Drut2009a} \cite{Drut2009b}, \cite{Armour2010}. Renormalization group calculations carried out perturbatively in the interaction strength  yield $\alpha^{\mathrm{c}}_{0} \simeq 0.833$ \cite{Vafek2008}. Analytics in the framework of the Schwinger-Dyson equation gives $\alpha^{\mathrm{c}}_{0} \simeq 1.13$ \cite{Khveshchenko2009}, and $\alpha^{\mathrm{c}}_{0} \simeq 1.62$ from the Bethe-Salpeter equation \cite{Gamayun2009}.

Taking into account the aforementioned asymptotics, we undertake a peeling-off procedure with the ansatz
\begin{equation}
\label{eqchiii}
\chi_4 = r^{\gamma_m - \tfrac{1}{2}} e^{-\mathrm{i} \tfrac{\varepsilon r}{2}} g(r),  
\end{equation}
so that Eq.~\eqref{eq634343} provides the following equation for $g(r)$: 
\begin{align}
\label{eqwewewe}
  &g''(r) + \left[ \tfrac{1}{r} \left( 2 \gamma_m + \tfrac{1}{1+\varepsilon r / \alpha} \right) - \mathrm{i} \varepsilon \right] g'(r) \nonumber \\
 &+ \bigg[ \tfrac{1}{r^2} \left( 1 - 2 \gamma_m \right) + \tfrac{\varepsilon}{r} \left( \tfrac{\alpha}{2} - \mathrm{i} \left\{ \gamma-1/2 \right\} \right)   \nonumber \\
&+ \tfrac{1}{r} \left( 1 + \tfrac{1}{1+\varepsilon r / \alpha} \right) \left( \tfrac{\gamma_m-1/2}{r} - \tfrac{\mathrm{i} \varepsilon}{2} \right)  \bigg] g(r) = 0.
\end{align}
Making the natural change of variable $\xi = - \varepsilon r / \alpha$ leads to the standard form of the confluent Heun equation \cite{Ronveaux} \cite{DowningHeun}
\begin{equation}
\label{eqheun}
  g''(\xi) + \left[ \rho + \tfrac{\beta+1}{\xi} + \tfrac{\lambda+1}{\xi-1} \right] g'(\xi) + \left[ \tfrac{\mu}{\xi} + \tfrac{\nu}{\xi-1} \right] g(\xi) = 0,
\end{equation}
where the parameters $\{ \rho, \beta, \lambda, \mu, \nu\}$ are given by
\begin{subequations}
\label{qes8}
 \begin{align}
  \rho = \mathrm{i} \alpha,~~~\beta = 2\gamma_m,~~~\lambda = -2, \label{vit8}
 \\
    \mu = \mathrm{i} \alpha \left( \gamma_m + \tfrac{1}{2} \right) + \left( \gamma_m - \tfrac{1}{2} \right) - \tfrac{\alpha^2}{2}, \label{vit88}
\\
\nu = -\mathrm{i} \tfrac{\alpha}{2} - \left( \gamma_m - \tfrac{1}{2} \right). \label{vit888}
 \end{align}
\end{subequations}
With the more familiar parameter parameterization $(\mu, \nu) \to (\delta, \eta)$ \cite{Note1}, we obtain
\begin{equation}
\label{qes559}
  \delta = - \tfrac{\alpha^2}{2},~~~~~~\eta = \tfrac{\alpha^2}{2} + \tfrac{3}{2}.
\end{equation}
The equation~\eqref{eqheun} has as its the solution the confluent Heun function $H_{\mathrm{C}} \left( \rho, \beta, \lambda, \delta, \eta, \xi \right)$ \cite{Ronveaux}, or explicitly
\begin{equation}
\label{qes9}
 g(r) = H_{\mathrm{C}} \left( \mathrm{i} \alpha, 2\gamma_m, -2, - \tfrac{\alpha^2}{2}, \tfrac{\alpha^2}{2} + \tfrac{3}{2}, - \tfrac{\varepsilon r}{ \alpha} \right),
\end{equation}
which, with Eq.~\eqref{eqchiii}, completes the formal solution of $\chi_4$. Therefore, with Eqs.~\eqref{eqasasasas3} we obtain the full solution of the ultrarelativistic Dirac-Kepler two-body problem, which reads
\begin{equation}
\label{eq2full}
\scalemath{0.9}{
\left(
 \begin{array}{c}
\chi_1 \\
 \chi_2 \\
 \chi_3 \\
\chi_4
 \end{array}
\right) = C r^{\gamma_m - \tfrac{1}{2}} e^{-\mathrm{i} \tfrac{\varepsilon r}{2}} 
\left(
 \begin{array}{c}
- \tfrac{2 m}{r} \tfrac{1}{\varepsilon + \alpha/r} g(r) \\
 - \tfrac{2}{r} \tfrac{1}{\varepsilon + \alpha/r} \left[ g'(r) + \left( \tfrac{1}{r} - \tfrac{\mathrm{i}\varepsilon}{2} \right) g(r) \right] \\
 0 \\
g(r)
 \end{array}
\right)},
\end{equation}
where $C$ is a normalization constant. Further analysis of the exact solution of Eq.~\eqref{eq2full}, using relations for the asymptotics of the confluent Heun function, promises to explain the nature of two-body quasibound states, charge criticality and condensation in Dirac materials \cite{Kotov2012}. Additionally, calculations on the two-body problem beyond the pure Coulomb interaction, to take into account dielectric and even dynamical screening, is an important open problem.

The coupling of two massless Dirac fermions into a truly bound state remains an outstanding and nontrivial task. As can be expected from the form of the principle two-particle Eq.~\eqref{eq634343}, which is highly reminiscent of the key one-particle Eq.~\eqref{eq6}, one is forced to consider zero-energy states in a bid to bind ultra-relativistic particles, which we now turn to. 


\subsection{Coupling at zero-energy} Now we consider zero-energy states in a faster than Coulomb potential at zero-energy ($\varepsilon = 0$). We choose to work with the Lorentzian interaction of Eq.~\eqref{toy1} to gain an analytic understanding of the system \cite{Portnoi2017}, and we note that the sign of the interaction is irrelevant for the quantization condition derived from Eq.~\eqref{eq634343}, since at zero-energy the interaction only enters as square or as a logarithmic derivative.

When $r \sim 0$, one finds the typical short-range behavior $\chi_4 \sim r^{|m|}$. Meanwhile, the asymptotic behavior as $r \to \infty$ is given by the decay $\chi_4 \sim r^{|m|-2 l_m}$, where
\begin{equation}
\label{eq121212}
l_m = \tfrac{1}{2} \left( 1+|m|+\sqrt{m^2+1} \right),
\end{equation}
is a key dimensionless quantity. Thus, we are motivated to seek a solution of Eq.~\eqref{eq634343} with the following ansatz and variable change
\begin{equation}
\label{eq1217878212}
\chi_4 = \frac{\xi^{\tfrac{|m|}{2}}}{(1+\xi)^{l_m}} f(\xi), \quad \xi = (r/d)^2.
\end{equation}
Here $f(\xi)$ is a polynomial in $\xi$ that does not affect the short- and long-range behavior of $\chi_4$ and it is determinable from Eq.~\eqref{eq634343} via:
\begin{align}
\label{eq121213}
 &\xi(1+\xi)^2 f''(\xi) \nonumber \\
 &+ (1+\xi) \left[ 1 + |m| + (2+|m|-2 l_m) \xi \right] f'(\xi) \nonumber \\
 &+ \left[ (\tfrac{U_0 d}{4})^2 - l_m^2 \right] f(\xi) = 0,
\end{align}
which is a form of the Gauss hypergeometric equation. Its solution, regular at $\xi = 0$, is given by
\begin{equation}
\label{eq11214}
 f(\xi) =  \,_2F_1\left(  l_m + \tfrac{1}{4} U_0 d  , l_m - \tfrac{1}{4} U_0 d ;  1 + |m| ; \tfrac{\xi}{\xi+1} \right).
\end{equation}
To ensure decaying solutions at infinity, one needs to terminate the power series in Eq.~\eqref{eq11214}. Then the radial wavefunction asymptotics of the full wavefunction are
\begin{equation}
\label{toy121232323232}
\lim_{r \to \infty}
\left(
 \begin{array}{c}
\chi_1 \\ \chi_2 \\ \chi_3 \\ \chi_4
 \end{array}
\right) \propto \frac{1}{r^{\sqrt{1+m^2}}} \left(
 \begin{array}{c}
m \\ 1 \\0 \\ \frac{1}{r}
 \end{array}
\right), 
\end{equation}
so that states with $m=0$ are marginally non-square-integrable, due to the slowly decaying component $\chi_2 \propto 1/r$. Therefore truly bound two-particle states have a non-vanishing $m$, and are known as zero-energy vortices.

\begin{figure}[htbp]
\includegraphics[width=0.5\textwidth]{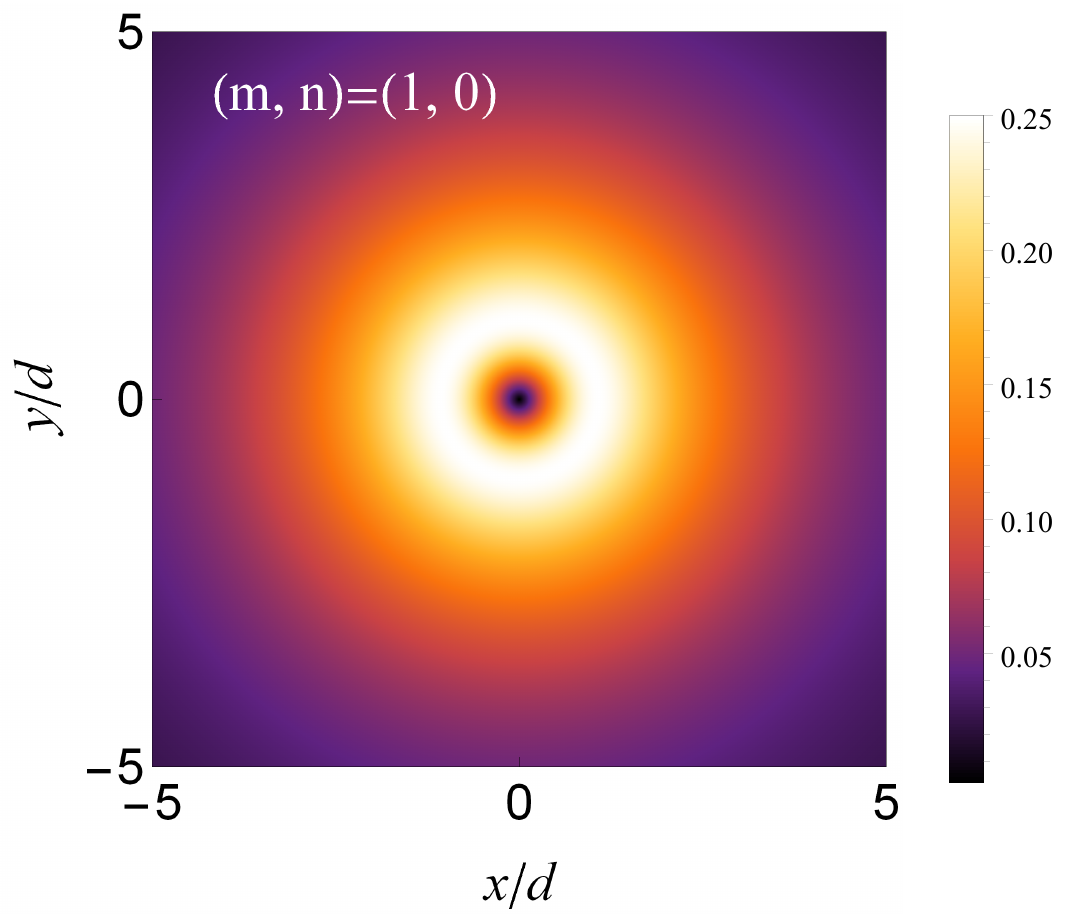}
\caption{ A plot of the radial probability density [see the color bar, right] of the lowest square-integrable two-particle bound state, characterized by the quantum numbers $(m, n) = (1, 0)$, as a function of position (in units of the length scale $d$).}
\label{figure1}
\end{figure}

The quantization condition for the formation of bound pairs follows from the second argument of Eq.~\eqref{eq11214} as
\begin{equation}
\label{eq11215}
 U_0 d = 4 (n + l_m), \quad n=0,1,2...
\end{equation}
where $l_m$ is defined in Eq.~\eqref{eq121212}. The marginally non-square-integrable states (with $m=0$) appear at values of $U_0d$ divisible by four [$U_0 d = 4 (n + 1)$], while the bound states (with $m\ne0$) occur at irrational values, starting from the threshold $U_0 d \simeq 6.83$. We plot in Fig.~\ref{figure1} the lowest square-integrable two-particle bound state, characterized by the quantum numbers $(m, n) = (1, 0)$, showing its vortex-like structure and axial symmetry.

This beautiful result of pairing of massless Dirac particles has some interesting implications \cite{Portnoi2017}. Namely, we see that a pair of ultrarelativistic fermions can form a bound state at zero-energy, if the interaction decays faster than the Coulomb interaction. Since the binding is independent of the sign of the inter-particle interaction, it suggests electron-electron pairing into bielectrons is as possible as electron-hole coupling into excitons. Furthermore, the obeyed bosonic statistics allows for a novel type of condensation, whose characteristics require further study.


\section{Confinement by other means}
\label{sec:other}

Here we briefly mention other proposals to bind massless Dirac particles, without recourse to zero-energy states, see also the review of Ref.~\cite{Rozhkov2011}.

The most popular method is to apply an external magnetic field perpendicular to the 2D material plane \cite{Martino2007}, \cite{Goerbig2011}. It was shown theoretically that various magnetic quantum dots are able to sustain either quasibound \cite{Ramezani2009} or fully bound states \cite{Roy2012}, \cite{Downing2016}, \cite{Downing2016b} depending on the spatial configuration. Fictitious or pseudo-magnetic fields, where certain strain configurations give rise to effective magnetic fields and even a type of quantum Hall effect, have also been extensively studied \cite{Guinea2010}, \cite{Guinea2010b}, \cite{Kim2011}.

It was also suggested theoretically that a spatially inhomogeneous Fermi velocity could lead to bound states \cite{Peres2009}, \cite{Concha2010}, \cite{Raoux2010}, \cite{Downing2017b}, a situation that might occur due to strain or perhaps superlattice effects. Other schemes have focused on opening up a band gap via substrate engineering \cite{Zhou2007}, or functionalizing the material to change its structure, for example the synthesis of fluorinated graphene \cite{Withers2010}.


\section{Zero modes in physics}
\label{sec:zeromodes}

We note in passing that zero-energy states are of great importance across several subfields of physics.

The inherent mathematical interest in zero-modes in non-relativistic quantum mechanics has been studied by Makowski and co-workers in Refs.~\cite{Makowski2007}, \cite{Makowski2007b}, \cite{Makowski2015}. In scattering theory, bound states at zero-energy are intrinsically linked to scattering phase shifts via the Levinson theorem, which has been investigated with the variable phase method in a series of papers \cite{Portnoi1988}, \cite{Portnoi1997}, \cite{Portnoi1998}, \cite{Portnoi1999}.

It has been shown theoretically that zero energy states in superconducting vortices can be fixed by particle-hole symmetry and accomplished with the aid of local lattice distortions \cite{Kusmartsev1998}. Such lattice deformations have been observed in a spin-polarized neutron scattering experiment on superconducting niobium \cite{Neumann1998}.

Zero-modes of Dirac equations are highly exotic \cite{Jackiw1981}, \cite{Brey2009}, since they can be associated with fractional charge \cite{Jackiw1984}, \cite{Jackiw2007}, topologically protected edge and soliton-like states \cite{Jackiw1976}, \cite{Ryu2002}, and celebrated Majorana modes \cite{Jackiw2012}. Most recently, analogues of some of these zero-energy state phenomena with Dirac-like equations have been both predicted and observed in a diverse array of systems, from photonics \cite{Kitagawa2012} to plasmonics \cite{Downing2018} to cold atoms \cite{Tewari2007}.


\section{Conclusions}
\label{sec:conc}

In this short review, we have described the problem of creating bound states in Dirac materials. The principle solution to the problem discussed here was the electrostatic confinement of zero-energy states. We have discussed some simple toy models of confinement at the single-particle level, which unveils the unique characteristics of bound states of massless Dirac particles. We have noted that the latest experimental literature displays signatures of zero-energy states in their reported data.

We have also described the ultra-relativistic two-body problem, and provided the solutions for the Dirac-Kepler problem and for the prototypical model of two-particle pairing. In particular, we have described how electron-electron pairing into bielectrons is remarkably both possible and energetically favorable. As a by-product of research into this topic, the number of analytic solutions of Dirac-like equations has been unexpectedly increased \cite{Bagrov2014}, showcasing how fundamental theory and applied physics are entwined in Dirac material research. 


\begin{acknowledgement}

This work was supported by the Government of the Russian Federation through the ITMO Fellowship and Professorship Program. C.A.D. acknowledges support from the Juan de la Cierva program (MINECO, Spain). M.E.P. acknowledges support from the EU H2020 RISE project CoExAN (Grant No. H2020-644076). We grateful to K.~S.~Gupta for fruitful discussions and C.A.D. thanks the AVE 03093 from Puerta de Atocha for hospitality during the writing of this paper. 

\end{acknowledgement}

\end{document}